\documentclass{PoS}
\usepackage{amssymb}
\usepackage{amsmath}
\usepackage{cite}
\usepackage{epsfig}

\usepackage{cite}
\usepackage{epsfig}

\newcommand\beq{\begin{equation}}
\newcommand\eeq{\end{equation}}
\newcommand\bea{\begin{eqnarray}}
\newcommand\eea{\end{eqnarray}}

\title{{\footnotesize DESY 08-106\\ SFB-CPP-08/100}\\
\boldmath Two-loop massive fermionic operator matrix elements and intial state QED 
corrections
to \boldmath{$e^+e^- \rightarrow \gamma^*/Z^*$}}

\ShortTitle{Two-loop massive fermionic operator matrix elements} 

\author{J. Bl\"umlein,$^a$ \speaker{A. De Freitas}$^{ab}$
and W. van Neerven$^c$~\footnote{Deceased.}\\
\llap{$^a$}DESY, Zeuthen, Platanenalle 6, D-15735 Zeuthen, Germany.\\
\llap{$^b$}Departamento de F\'isica, Universidad Sim\'on Bol\'ivar,
Apartado Postal 89000,\\
Caracas 1080-A,Venezuela.\\
\llap{$^c$}Institut-Lorentz, Universiteit Leiden,
P.O. Box 9506, 2300 HA Leiden, The Netherlands.}

\abstract{We describe the calculation of the two--loop massive operator matrix
  elements for  massive external fermions. These matrix elements are needed for the
  calculation of the $O(\alpha^2)$ initial state radiative corrections to $e^+e^-$ 
  annihilation into a neutral virtual gauge boson, based on the renormalization group
  technique.}

\FullConference{8th International symposium on radiative corrections, Florence, Italy,
October, 1-5 2007}

\begin{document}

%%%%%%%%%%%%%%%%%%%%%%%%%%%%%%%%%%%%%%%%%%%%%%%%%%%%%%%%%%%%%%%%%%%%%%%%%%%%%%%%%
\section{Introduction}
%%%%%%%%%%%%%%%%%%%%%%%%%%%%%%%%%%%%%%%%%%%%%%%%%%%%%%%%%%%%%%%%%%%%%%%%%%%%%%%%%

The $O(\alpha^2)$ initial state QED corrections to $e^+ e^-$ annihilation 
into a virtual photon or $Z$--boson were only calculated once in 
Ref.~\cite{BBN} in a standard Feynman diagram calculation. Due to the smallness
of the ratios $\xi = m^2/s'$, with $m$ the electron mass and $s'$ the cms momentum squared of the 
virtual gauge boson, power corrections
in $\xi$ are negligible and only the logarithmic and constant terms in this parameter
have to be maintained. The renormalization group equation, applied to this process, 
allows to derive all contributing terms in the on-mass-shell scheme. Our goal is to 
calculate all these terms and
to check the result derived in \cite{BBN} previously.

In this limit the differential scattering cross section is given in Mellin space by
\begin{eqnarray} 
\label{eqMA1}
\frac{d\sigma_{e^+e^-}}{ds'} &=&
\frac{1}{s} \sigma^{(0)}(s') 
   \Biggl\{
   1 + a_0 \left[ P_{ee}^{(0)} \cdot {\bf L}
   +\left(\tilde{\sigma}^{(0)}_{ee} + 2 \Gamma_{ee}^{(0)}\right)\right] 
%   \nonumber\\ 
%& &
   + a_0^2\Biggl\{\Biggl[
   \frac{1}{2} {P_{ee}^{(0)}}^2 - \frac{\beta_0}{2} P_{ee}^{(0)} + \frac{1}{4}
   P_{e \gamma}^{(0)} \cdot P_{\gamma e}^{(0)} \Biggr] {\bf L}^2
   \nonumber\\
& & %\hspace{7mm}
   +\Biggl[P_{ee}^{(1)} + P_{ee}^{(0)} \left( \tilde{\sigma}_{ee}^{(0)} + 
2 \Gamma_{ee}^{(0)}\right)
   - \beta_0 \tilde{\sigma}_{ee}^{(0)} +P_{\gamma e}^{(0)} 
\tilde{\sigma}_{e 
\gamma}^{(0)}
   + \Gamma_{\gamma e}^{(0)} P_{e \gamma}^{(0)} \Biggr] {\bf L}
   \nonumber\\
& & %\hspace{7mm} 
   + \left(2 \Gamma_{ee}^{(1)} + \tilde{\sigma}_{ee}^{(1)}\right) + 2 
\Gamma_{ee}^{0)} 
\tilde{\sigma}_{ee}^{(0)} + 2 \tilde{\sigma}_{e \gamma}^{(0)} 
\Gamma_{\gamma 
e}^{(0)}
+ {\Gamma_{ee}^{(0)}}^2 \Biggr\} \Biggr\}  \,\, .
\end{eqnarray} 
Here the $\Gamma$'s denote the massive operator matrix elements, the
$\tilde{\sigma}$'s are massless Wilson coefficients, and the $P$'s are splitting
functions. $\sigma^{(0)}(s)$ is the
Born level scattering cross section in terms of the invariant mass $s$ of the
initial $e^+e^-$ pair and ${\bf L} = \ln\left(s'/m^2 \right)$.
It is convenient to represent the differential scattering cross section
in terms of three contributions, the flavor non-singlet terms with a 
single fermion line (I), those with a closed fermion line (II), and the
pure-singlet terms (III).~\footnote{In Ref.~\cite{BBN} four contributions 
were considered dividing those to I into two pieces according to the
genuine $2 \rightarrow 3$ particle scattering cross sections.}
These contributions are~:
%----------------------------------------------------------------------------- 
\begin{eqnarray} 
\label{eqMA1a}
\frac{d\sigma_{e^+e^-}^{\rm I}}{ds'} &=&
\frac{1}{s} \sigma^{(0)}(s') 
   \Biggl\{
   1 + a_0 \left[ P_{ee}^{(0)} \cdot {\bf L}
   +\left(\tilde{\sigma}^{(0)}_{ee} + 2 \Gamma_{ee}^{(0)}\right)\right] 
   + a_0^2\Biggl\{
   \frac{1}{2} {P_{ee}^{(0)}}^2 {\bf L}^2
   \nonumber\\
& & %\hspace{21mm}
   +\Biggl[P_{ee}^{(1),{\rm I}} 
   + P_{ee}^{(0)} \left( \tilde{\sigma}_{ee}^{(0)} + 
   2 \Gamma_{ee}^{(0)}\right) \Biggr]{\bf L}
%   \nonumber\\
%& & \hspace{21mm} 
   + \left(2 \Gamma_{ee}^{(1),{\rm I}} + 
\tilde{\sigma}_{ee}^{(1),{\rm I}}\right) + 2 
\Gamma_{ee}^{0)} 
\tilde{\sigma}_{ee}^{(0)} 
+ {\Gamma_{ee}^{(0)}}^2 \Biggr\} \Biggr\}  
\\
\label{eqMA1b}
\frac{d\sigma_{e^+e^-}^{\rm II}}{ds'} &=&
\frac{1}{s} \sigma^{(0)}(s') 
   a_0^2\Biggl\{- \frac{\beta_0}{2} P_{ee}^{(0)} {\bf L}^2
   +\Biggl[P_{ee}^{(1), {\rm II}} - \beta_0 \tilde{\sigma}_{ee}^{(0)} 
     \Biggr] {\bf L}
   + \left(2 \Gamma_{ee}^{(1),{\rm II}} + 
\tilde{\sigma}_{ee}^{(1),{\rm II}}\right) \Biggr\}   
\\
\label{eqMA1c}
\frac{d\sigma_{e^+e^-}^{\rm III}}{ds'} &=&
\frac{1}{s} \sigma^{(0)}(s') 
   a_0^2\Biggl\{\frac{1}{4}
   P_{e \gamma}^{(0)} \cdot P_{\gamma e}^{(0)} {\bf L}^2
   +\Biggl[P_{ee}^{(1),{\rm III}} + P_{\gamma e}^{(0)} 
\tilde{\sigma}_{e 
\gamma}^{(0)}
   + \Gamma_{\gamma e}^{(0)} P_{e \gamma}^{(0)} \Biggr] {\bf L}
   \nonumber\\
& & \hspace{21mm} 
   + \left( 2 \Gamma_{ee}^{(1),{\rm III}} + \tilde{\sigma}_{ee}^{(1),{\rm 
III}}\right)+ 2 \tilde{\sigma}_{e \gamma}^{(0)} 
   \Gamma_{\gamma e}^{(0)} 
\Biggr\}~.   
\end{eqnarray} 
%-----------------------------------------------------------------------------

The massless Wilson coefficients are known in the literature in the
$\overline{\rm MS}$ scheme for the Drell--Yan process \cite{DY1,DY2}.
The leading and next-to-leading order splitting functions $P^{(0)}_{ij}$ and
$P^{(1)}_{ij}$ are well-known from QCD. The only missing piece we need
to calculate are the massive fermionic operator matrix elements 
$\Gamma_{ij}^{(k)}$ to $O(\alpha^2)$. In
the following sections, we describe this calculation.

%%%%%%%%%%%%%%%%%%%%%%%%%%%%%%%%%%%%%%%%%%%%%%%%%%%%%%%%%%%%%%%%%%%%%%%%
\section{Calculation techniques}
%%%%%%%%%%%%%%%%%%%%%%%%%%%%%%%%%%%%%%%%%%%%%%%%%%%%%%%%%%%%%%%%%%%%%%%%

\begin{figure}

\begin{minipage}[h]{1.0\linewidth}
\centering\epsfig{figure=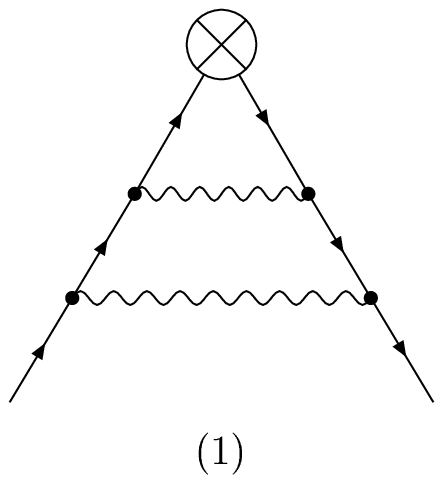,width=0.14\linewidth} \hspace{1mm}
\centering\epsfig{figure=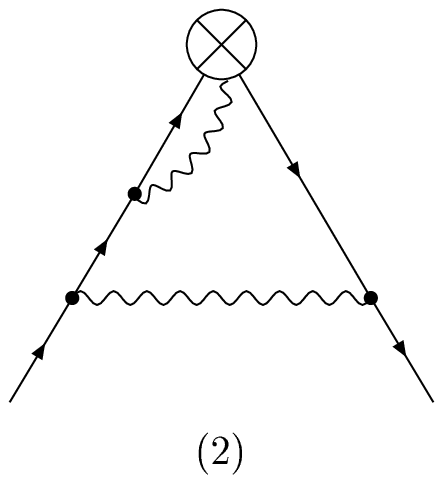,width=0.14\linewidth} \hspace{1mm}
\centering\epsfig{figure=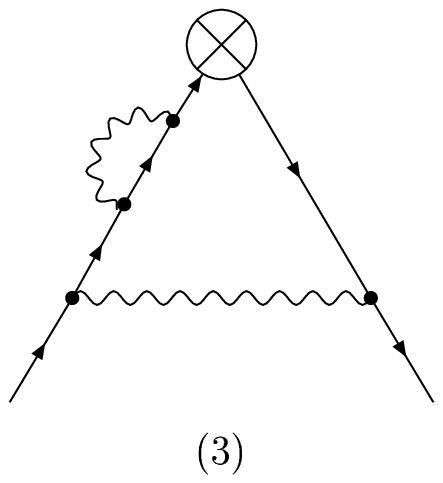,width=0.14\linewidth} \hspace{1mm}
\centering\epsfig{figure=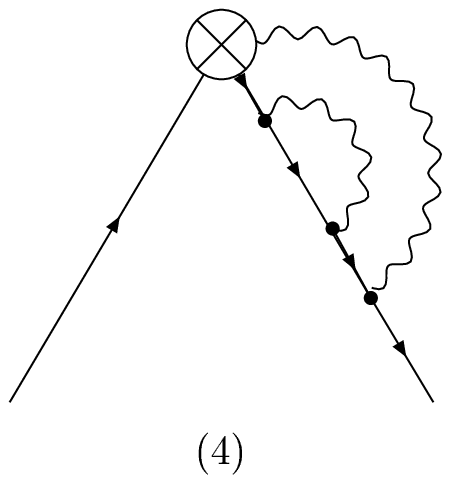,width=0.14\linewidth} \hspace{1mm}
\centering\epsfig{figure=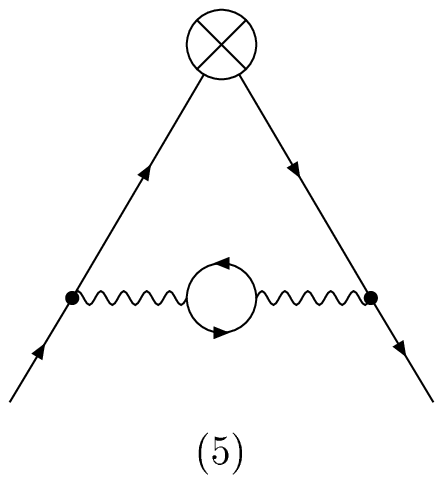,width=0.14\linewidth} \hspace{1mm}
\centering\epsfig{figure=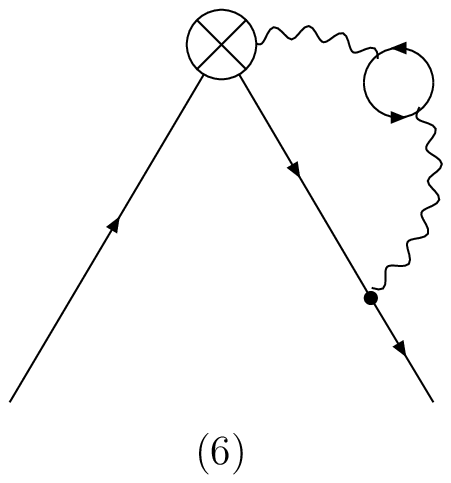,width=0.14\linewidth} \hspace{1mm}

\vspace{2mm}
\centering\epsfig{figure=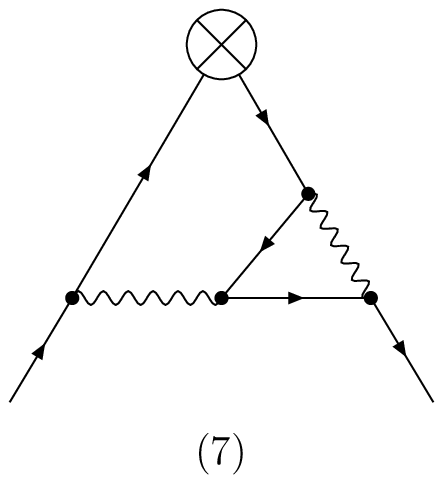,width=0.14\linewidth} \hspace{1mm}
\centering\epsfig{figure=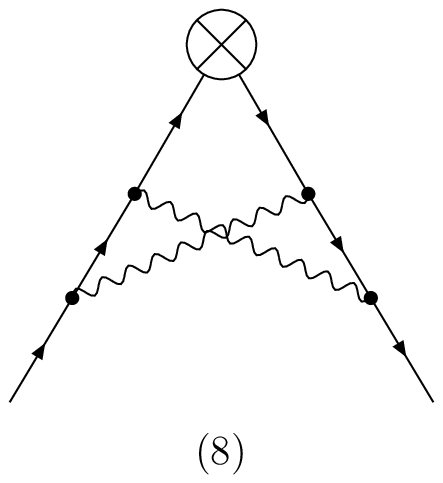,width=0.14\linewidth} \hspace{1mm}
\centering\epsfig{figure=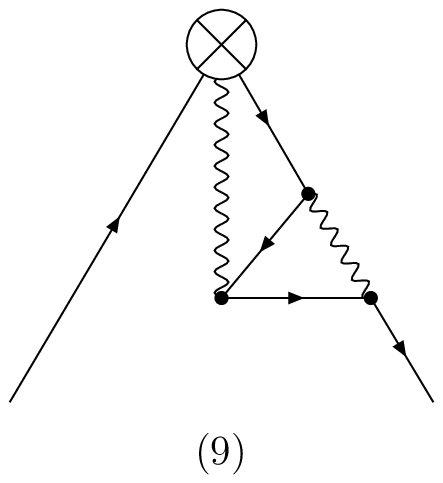,width=0.14\linewidth} \hspace{1mm}
\centering\epsfig{figure=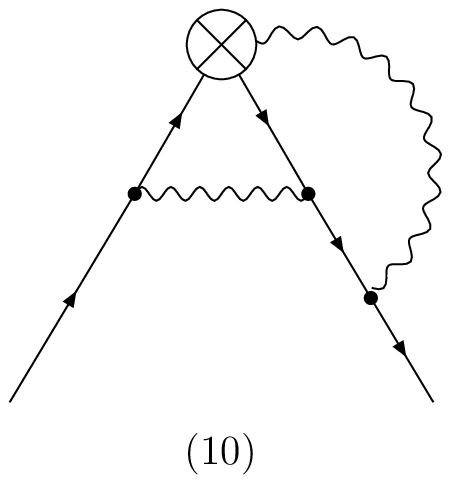,width=0.14\linewidth} \hspace{1mm}
\centering\epsfig{figure=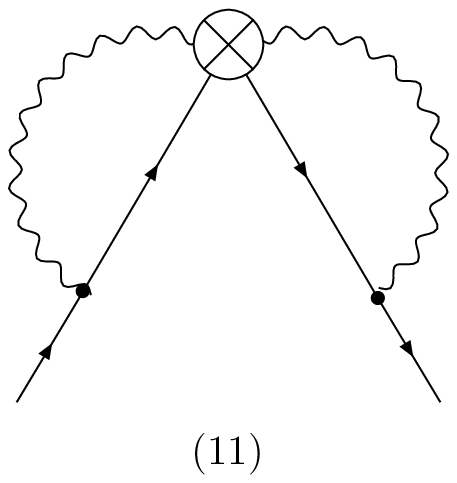,width=0.14\linewidth} \hspace{1mm}
\centering\epsfig{figure=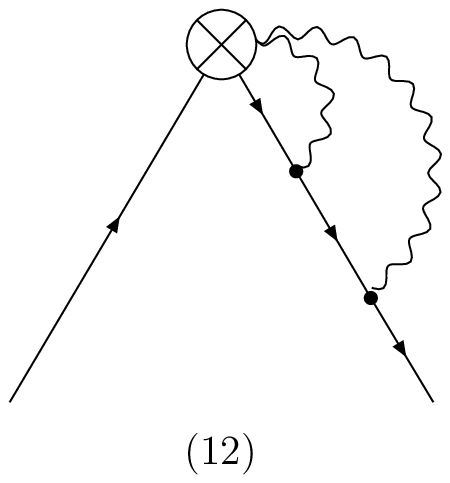,width=0.14\linewidth} \hspace{1mm}
\end{minipage}

\caption{Two-loop diagrams contributing to the massive operator matrix
  elements. The antisymmetric diagrams count twice.}
\label{fig1}
\end{figure}

\begin{figure}
\centering\epsfig{figure=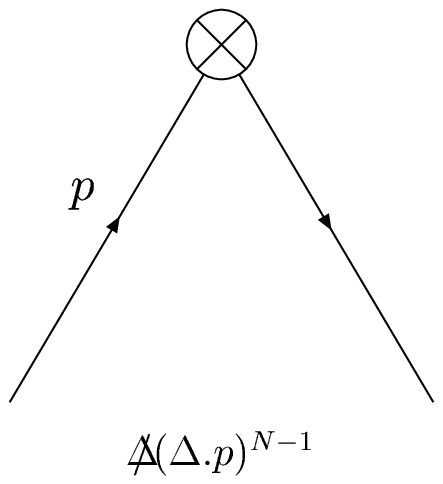,width=0.15\linewidth} \hspace{1mm}
\centering\epsfig{figure=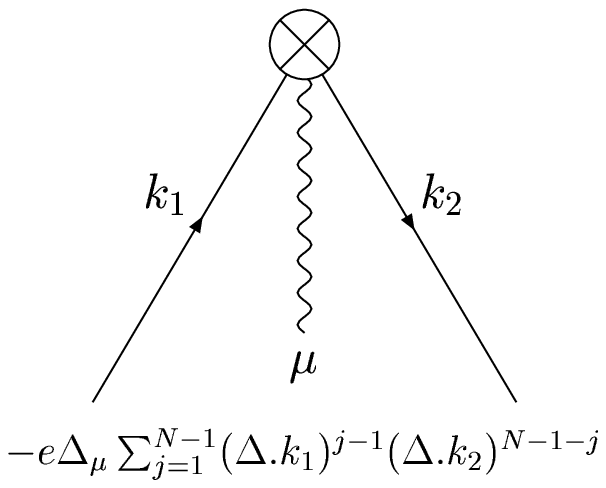,width=0.22\linewidth} \hspace{1mm}
\centering\epsfig{figure=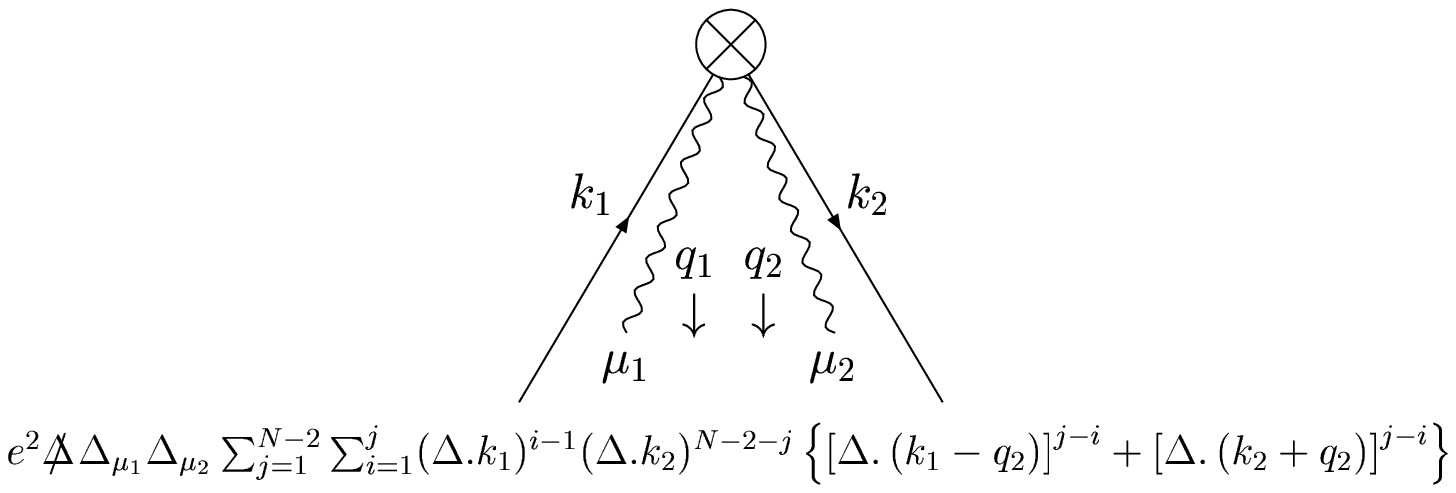,width=0.54\linewidth} \hspace{1mm}
\caption{Feynman rules.}
\label{fig2}
\end{figure}

To obtain $\Gamma_{ee}^{(1), \rm I-III}$ we calculate the 
diagrams in figure \ref{fig1} (and the corresponding self-energy diagrams) 
using the rules shown in figure \ref{fig2} for an external massive fermion 
(anti-fermion)~\footnote{We recalculated the LO OME's and corrected typographical 
errors in \cite{BBN}.} for
general integer values of the Mellin-moment $n$ in the form
%---------------------------------------------------------------------------
\begin{eqnarray}
\widetilde{\Gamma}_{ee}^{(1)}(n) 
= \int_0^1 d z z^{n-1} \Gamma_{ee}^{(1)}(z)~.
\end{eqnarray}
%---------------------------------------------------------------------------
In this way we can separate the splitting function $P_A(x)$ \cite{CFP} from the 
remainder non--singlet splitting function in the single pole term 
$1/\varepsilon$ and in the constant part of the operator matrix element.

Since we consider the radiative corrections to the time-like 
process~\footnote{In space-like processes as deeply inelastic scattering
$e^- \rightarrow e^+$ transitions do even occur in $O(\alpha^2 L^2)$, 
cf.~\cite{R6}.} $e^+ e^- \rightarrow Z_0$
up to two-loop order for both initial states, the particle nature
has to be conserved, i.e. only $e^- \rightarrow e^-$ and  $e^+ \rightarrow e^+$
transitions contribute, which requires a corresponding projection in
$\Gamma_{ee}^{(1)}(x)$.

All of the diagrams can be written as a linear combination of integrals of the 
following type:
\begin{eqnarray}
A^{a,b}_{\nu_1,\nu_2,\nu_3,\nu_4,\nu_5} &=&
\int \frac{d^D k_1}{(4\pi)^D} \frac{d^D k_2}{(4\pi)^D} 
\frac{ (\Delta \cdot k_1)^{a} (\Delta \cdot k_2)^{b}
}{D_1^{\nu_1}D_2^{\nu_2}D_3^{\nu_3}D_4^{\nu_4}D_5^{\nu_5}}
\\
B^{a,b}_{\nu_1,\nu_2,\nu_3,\nu_4,\nu_5} &=&
\int \frac{d^D k_1}{(4\pi)^D} \frac{d^D k_2}{(4\pi)^D} 
\frac{ k_2 \cdot p (\Delta \cdot k_1)^{a} (\Delta \cdot k_2)^{b}
}{D_1^{\nu_1}D_2^{\nu_2}D_3^{\nu_3}D_4^{\nu_4}D_5^{\nu_5}}
\\
F^{a,b}_{\nu_1,\nu_2,\nu_3,\nu_4,\nu_5} &=&
\int \frac{d^D k_1}{(4\pi)^D} \frac{d^D k_2}{(4\pi)^D} 
\frac{ (\Delta \cdot k_1)^{a} (\Delta \cdot k_2)^{b} 
}{D_1^{\nu_1}D_2^{\nu_2}D_3^{\nu_3}D_4^{\nu_4}D_5^{\nu_5}}
\sum_{j=0}^{n-1} (\Delta \cdot p)^j (\Delta \cdot k_1)^{n-1-j} 
\\
G^{a,b}_{\nu_1,\nu_2,\nu_3,\nu_4,\nu_5} &=&
\int \frac{d^D k_1}{(4\pi)^D} \frac{d^D k_2}{(4\pi)^D} 
\frac{ (\Delta \cdot k_1)^{a} (\Delta \cdot k_2)^{b} 
}{D_1^{\nu_1}D_2^{\nu_2}D_3^{\nu_3}D_4^{\nu_4}D_5^{\nu_5}}
\sum_{j=0}^{n-1} (\Delta \cdot k_1)^j (\Delta \cdot k_2)^{n-1-j}
\end{eqnarray}
and a couple more of integrals associated with the diagrams that have two
photons coming out of the vertex. Here $D_1 = k_1^2-m^2$, $D_2 = k_2^2-m^2$,
$D_3 = (k_1-p)^2$, $D_4 = (k_1-k_2)^2$ and $D_5 = (k_2-k_1+p)^2-m^2$, with $p$
the momentum of the electron. For
example, the cross-box diagram appearing in figure \ref{fig1} can be
written in terms of these integrals as follows
\begin{eqnarray}
&&
\frac{1}{2} (D-2) (D-4) \left[
2 A^{0,1 + n}_{02110} - 2 A^{0,1 + n}_{12001} + A^{1,n}_{01111}
- A^{1,n}_{11011} + A^{1,n}_{11101} - A^{1,n}_{11110} - (\Delta \cdot  
p) A^{0,n}_{01111}
\right]
\nonumber \\
&&
- \frac{1}{2} (D-2) (D-8) \left[
A^{0,1 + n}_{01111} - A^{0,1 + n}_{11011}
+ (\Delta \cdot p) A^{0,n}_{11101} - (\Delta \cdot p) A^{0,n}_{10111}
\right]
-16 m^4 A^{0,1 + n}_{12111}
\nonumber \\
&&
+ 2 (D-2) \left[
- A^{0,1 + n}_{11101} + A^{0,1 + n}_{11110} + (\Delta \cdot p) A^{0,n}_{11011}
\right]
+ 2 (D-4) m^2 \left[
A^{0,1 + n}_{11111} - (\Delta \cdot p) A^{0,n}_{11111}
\right]
\nonumber \\
&&
+8 m^2 A^{0,1 + n}_{12011}
+8 m^2 A^{0,1 + n}_{12101}
-4 m^2 A^{0,1 + n}_{12110}
-4 m^2 A^{0,1 + n}_{02111}
-4 (D-2) \left[
B^{0,1 + n}_{12011} + B^{0,1 + n}_{12101}
\right]
\end{eqnarray}

These integrals can be calculated in various ways.
In case of external gluon lines the $O(\alpha^2_s)$  the integrals appearing
in the massive operator matrix
elements \cite{BBK} can be calculated using Mellin-Barnes techniques and
representations in terms of generalized hypergeometric functions without
using the integration-by-parts technique. Working in Mellin space one may 
solve difference equations for harmonic sums \cite{HSUM}, cf. \cite{BBK2}, 
which leads
to ab initio compact expressions. In the present calculation we follow the
more traditional way of calculation, which was also used in \cite{BUZA}.
The massive 5--propagator integrals are reduced to 4--propagator
integrals using the integration-by-parts technique \cite{GAUSS}. 
For example, one of the most complicated integrals coming from the cross-box
diagram, namely $A^{0,n}_{12111}$, can be decomposed as follows
\begin{eqnarray}
A^{0,n}_{12111} &=& \frac{1}{\epsilon} 
                     \left[ \phantom{\frac{|}{|}} 
                     -A^{0,n}_{12210}-A^{0,n}_{02211}-A^{0,n}_{12102}-A^{0,n}_{22101} 
                     \right. 
                     \nonumber \\ && 
                      \phantom{ \frac{1}{\epsilon} \left[ \right. }
                     +\frac{1}{1+\epsilon}
                     \left(
                     -2A^{0,n}_{31101} + 2A^{0,n}_{30111} - A^{0,n}_{21210} 
                     +2A^{0,n}_{20211} - A^{0,n}_{21102}
                     \right.
                     \nonumber \\  && 
                     \phantom{ \frac{1}{\epsilon} \left[ \right. +\frac{1}{1+\epsilon} \left(
                     \right. }
                     \left. \left.
                     - A^{0,n}_{21201} - 2A^{0,n}_{11310} - 2A^{0,n}_{10311} 
                     - A^{0,n}_{01311} -A^{0,n}_{11202} \right) \phantom{\frac{|}{|}} \right]~.
\end{eqnarray}

The 4--propagator integrals obey
representations in terms of up to three Feynman parameter integrals over 
the unit cube. In course of the calculation we arrange the structure of 
the Feynman parameter integrals such that they have the form 
%------------------------------------------------------------------------
\begin{eqnarray}
\label{Fint}
F(\varepsilon,N) = \int_0^1 dx~x^N~~\int_0^1 dy \int_0^1 dz~f(x,y,z; 
\varepsilon)~.
\end{eqnarray}
%------------------------------------------------------------------------
The arts of integration consist now in performing the numerous integrals 
(\ref{Fint}) in Feynman parameter space. This partly requires 
Mellin-Barnes representations, repeated conformal mapping of integration 
variables, the use of parameter mappings for hypergeometric integrals,
and extended regularization. Numerous integrals of (Nielsen) 
polylogarithms  \cite{NIELS} normally of complicated arguments have to be 
performed.  It turns out that the final results can be expressed in terms
of Nielsen integrals, partly weighted by denominators $1/x, 1/(1 - x)^k,
1/(1+x)^l,~~k \leq 3, l \leq 2$. This structure is expected similar to
that occurring in case of a wide variety of other 2--loop quantities
\cite{MATH}. For example, the integral mentioned above
gives
\begin{eqnarray}
A^{0,n}_{12111} &=&
\int_0^1 dx \,\, x^n \left\{ \, 
\frac{2}{3} {\rm Li}_2(1-x) 
+\frac{1}{3} \ln^2(x)
+ \frac{2}{3(1-x)} \ln(x) 
+\frac{2}{3} \zeta_2 \right. \nonumber\\ 
&& \phantom{ \int_0^1 dx \,\, x^N \left\{ \right. }
-\frac{1}{3}(1-x)^{-3-2\epsilon} \ln^2(x)
-\frac{2}{3}(1-x)^{-2-2\epsilon} \ln(x) \nonumber\\
&& \phantom{ \int_0^1 dx \,\, x^N \left\{ \right. }
+\frac{2}{3}(1-x)^{-1-2\epsilon}
+\epsilon (1-x)^{-1-2\epsilon} \left( 2 \zeta_2 -1 \right) \nonumber\\
&& \phantom{ \int_0^1 dx \,\, x^N \left\{ \right. }
+(-1)^n \left[ \,
\frac{4}{3} {\rm Li}_2(-x) 
- \frac{2}{3} \left( 1 - \frac{2}{(1+x)^3} \right) {\rm Li}_2(1-x) \right. \nonumber\\
&& \phantom{ \int_0^1 dx \,\, x^N \left\{ \right. + (-1)^N \left[ \right. }
- \left( 1 - \frac{1}{(1+x)^3} \right) \ln^2(x)
+ \frac{4}{3} \ln(x) \ln(1+x) \nonumber\\
&& \phantom{ \int_0^1 dx \,\, x^N \left\{ \right. + (-1)^N \left[ \right. }
+ \frac{2}{3(1+x)} \left( \frac{2}{(1+x)^2} - \frac{1}{(1+x)} - 1 \right) 
\ln(x) \nonumber\\
&& \phantom{ \int_0^1 dx \,\, x^N \left\{ \right. + (-1)^N \left[ \right. }
\left. \left.
- \frac{2}{3(1+x)} + \frac{4}{3(1+x)^2} + \frac{2}{3} \zeta_2 \right] \right\}~.
\end{eqnarray}

\noindent
Detailed checks of the analytic calculations were performed using 
numerical integrations at high precision with {\tt MAPLE} for fixed 
moments and individual integrals whenever possible. Mellin-Barnes
integrals were performed for comparison in a series of fixed moments using the code {\tt MB} 
\cite{MB}. We also compared to a few  low (0th--2nd) moments for some of the
scalar integrals to the results obtained using  {\tt tarcer} \cite{TARCER} based on 
the
algorithm \cite{OT}.

We compared the results of our calculation to that in \cite{BBN} in the logarithmic 
orders and agree after correcting a series of typographical errors in some 
functions there,
decomposing the scattering cross section according to the renormalization
group. The final result including the constant term will be published soon.

\section{Conclusion}

Since the scale $m^2/s' \ll 1$ the QED initial state radiative corrections to
$e^+ e^- \rightarrow Z^*/\gamma^*$ to $O(\alpha^2)$ can be calculated neglecting the power 
corrections.
Alternatively to a direct calculation one may use the renormalization group to
arrange the calculation in terms of convolutions of splitting functions, massless
Wilson coefficients and process--independent massive operator matrix elements. The
latter quantities with outer massive fermion lines form the essential part of the 
present calculation. We used the integration-by-parts technique to reduce the 
integrals to 4--propagator integrals, which lead to Mellin moments in terms of
triple Feynman parameter integrals. The differential scattering cross section
can be obtained as assembly of the massive operator matrix elements, massless Wilson
coefficients for the Drell-Yan process and the splitting functions to two--loop 
orders. 

\vspace{2mm}
\noindent
{\bf Acknowledgement.}\\
This work was supported in part by the 
Alexander-von-Humboldt Foundation and by DFG Sonderforschungsbereich 
Transregio 9, Computergest\"utzte Theoretische Physik. For collaboration 
in an early phase of this paper we would like to thank A.~Mukherjee.

%%%%%%%%%%%%%%%%%%%%%%%%%%%%%%%%%%%%%%%%%%%%%%%%%%%%%%%%%%%%%%%%%%%%%%%%%

\end{document}